# Clustering and Uncertainty in Perfect Chaos Systems

Sergey A. Kamenshchikov*

*Moscow State University of M.V.Lomonosov, Faculty of Physics,*
*Russian Federation, Moscow, Leninskie Gory, Moscow, 119991*
*Corresponding e-mail: kamphys@gmail.com*

---

## Abstract

The goal of this investigation was to derive strictly new properties of chaotic systems and their mutual relations. The generalized Fokker-Planck equation with a non stationary diffusion has been derived and used for chaos analysis. An anomalous transport turned out to be natural property of this equation. A nonlinear dispersion of the considered motion allowed to find a principal consequence: a chaotic system with uniform dynamic properties tends to instable clustering. Small fluctuations of particles density increase by time and form attractors and stochastic islands even if the initial transport properties have uniform distribution. It was shown that an instability of phase trajectories leads to the nonlinear dispersion law and consequently to a space instability. A fixed boundary system was considered, using a standard Fokker-Planck equation. We have derived that such a type of dynamic systems has a discrete diffusive and energy spectra. It was shown that phase space diffusion is the only parameter that defines a dynamic accuracy in this case. The uncertainty relations have been obtained for conjugate phase space variables with account of transport properties. Given results can be used in the area of chaotic systems modelling and turbulence investigation.

---

## 1 Introduction

Several scenarios of a turbulence transition have been proposed since 1883 when the turbulence concept was firstly introduced by an English engineer Osborne Reynolds. The dynamic phase transition in a liquid stream was remarked by Reynolds – it was characterized by an unstable vortex_appearance and two limit states of motion: laminar and turbulent types. Since an introduction of a turbulence concept its properties have been generalized and transformed into properties of a chaos state.

Several scenarios of a turbulence transition have been developed. Among them the Landau – Hopf instability mechanism [1], the Lorenz attractor mechanism [2], the scenario of Poincare – Feigenbaum [3] and Kolmogorov - Arnold – Moser [4]. Each of the outlined mechanisms has its basic assumptions and an individual area of its application. For this reason none of them suggests some universal approach - moreover unambiguous connections between the given mechanisms are not stated yet. Let's try to highlight common points for the formulation of chaos concept.

Consent has been obtained that an unpredictability of chaos is consequence of two conditions: a finite resolution of generalized phase space, instability of phase trajectories and mixing of phase trajectories [5]. Formally these conditions can be defined in the following way – (1), (2) and (3):

$$_{\Delta}x^{i} \geq {}_{\Delta}x^{i}{}_{\min} \succ 0 \tag{1}$$

Here ${}_{\Delta}x^{i}{}_{\min}$ is an element of the Hilbert phase space while $\vec{x}(t)$ is the characteristic vector – each dynamic state corresponds to a given $\vec{x}$. For the case of independent agents a dimension of the phase space is defined according to a number of agents and Euclidean dimension.

*The author declares that there is no conflict of interests regarding the publication of this article.

A second condition can be expressed through the sum of positive Lyapunov factors $\sigma_i^+$ for an each dimension of phase space:

$$\sum_{i=1}^{P} \sigma_i^+(\vec{x}) \succ 0 \tag{2}$$

Here $P$ corresponds to a number of a phase space dimensions. A realization of these chaos conditions leads to another basic consequence. Under the condition of a phase space mixing reverse mapping for time $t(\vec{x})$ is not single valued in general case. Mixing property can be expressed through the auto correlation function $R\left(f(\vec{x}),\tau\right)$, where $f$ is an arbitrary dynamic function (for example momentum):

$$\lim_{\tau \to \infty} R\left(f(\vec{x}),\tau\right) = 0 \tag{3}$$

Here $\tau$ is a delay time between the start and the end of a system evolution observation. Condition (3) is naturally satisfied for random walks. A realization of this condition automatically leads to an execution of the Slutsky criterion for an ergodic system:

$$\lim_{\substack{T \to \infty \\ \tau \to \infty}} \int_{0}^{T} R\left(f(\vec{x}),\tau\right) \cdot \left(1 - \frac{\tau}{T}\right) d\tau = 0 \tag{4}$$

According to (4) system becomes ergodic for $\tau \to \infty$. For real physical systems ergodicity corresponds to the following relation:

$${}_{\Delta}t_{\min} \succ {}_{\Delta}t_{ins} \qquad {}_{\Delta}t_{inst} = \frac{1}{h} = \frac{1}{\sum_{i=1}^{K} \sigma_i^+(\vec{x})} \qquad \langle h \rangle = \bar{h} = h_d \tag{5}$$

Here ${}_{\Delta}t_{\min}$ is an available time resolution and ${}_{\Delta}t_{inst}$ is the instability increment expressed through the integrated Lyapunov factors. Andrey Kolmogorov has shown that an averaged $h$ corresponds to the Gibbs entropy production:

$$h_d = \frac{\partial S}{\partial t} = \frac{\partial}{\partial t} \ln\left(\Gamma\right) \tag{6}$$

In such a way positive $h$ necessarily leads to the positive dynamic entropy. Finally we may define complex portrait of chaotic system – it is a complex dynamic system that corresponds to given obligatory properties: a finite resolution of phase space, instability of phase trajectories, mixing of phase trajectories and a positive entropy production. In next sections we shall discuss relative and nonlinear properties of chaos in details. However all properties that have been shown above are assumed to be satisfied.

## 2 Nonlinear dispersion and clustering

Let's consider one of possible approaches to the description of chaos - the partial differential equation of Fokker-Planck. This model has been modified and finalized by Kolmogorov in 1938 [5]. The basis used for its formulation is Chapman - Kolmogorov equation. If we consider a one dimensional case then a transitional probability for random walks $x_1, t_1 \to x_3, t_3$ formally satisfies the following markovian relation:

$$W(x_3,t_3|x_1,t_1) = \int dx_2 W(x_3,t_3|x_2,t_2) \cdot W(x_2,t_2|x_1,t_1) \tag{7}$$

In the above formula *x*(*t*) is a system coordinate, while $W(x,t|x_0,t_0)$ is a probability density of $x(t)$ location subject to the initial coordinate is $x_0(t_0)$. Fokker – Planck – Kolmogorov (*FPK*) equation has been received based on following assumptions:

**A1**. $W(x',t'|x,t) = W(x',x,t'-t) = W(x',x,{}_\Delta t)$. A transitional probability doesn't depend on the initial time point. This demand implies that condition of $\tau \succ\succ \tau_C$ is satisfied, where $\tau_C$ is an effective width of auto correlation function for x(t);

**A2**. $P(x',t) = W(x',x,t)$. A final probability doesn't depend on the initial coordinate. This restriction implies $\tau \succ\succ \tau_c$ as well;

**A3** The initial distribution density is defined by Dirac delta function: $W(x) = \delta(0)$ - we assume that the initial coordinate can be defined quite accurately in relation to the general size of a considered system;

**A4.** According to A3 assumption an approximation of the second order may be received for the transitional probability:

$$W(x,x_0,{}_\Delta t) = \delta(x-x_0) + a(x,{}_\Delta t)\cdot\delta'(x-x_0) + \frac{1}{2}\cdot b(x,{}_\Delta t)\cdot\delta''(x-x_0) \qquad (8)$$

Factors $a(x,{}_\Delta t)$ and $b(x,{}_\Delta t)$ are defined by relations (8) and (9):

$$a(x,{}_\Delta t) = \int (x-x_0)\cdot W(x,x_0,{}_\Delta t)dx_0 = \langle\langle {}_\Delta x\rangle\rangle \qquad (9)$$

$$b(x,{}_\Delta t) = \int (x-x_0)^2\cdot W(x,x_0,{}_\Delta t)dx_0 = \langle\langle {}_\Delta x^2\rangle\rangle \qquad (10)$$

A substitution of expansion (8) into Chapman - Kolmogorov equation [6] gives us the following form for $P(x',t) = W(x',x,t)$ - standard Fokker-Planck equation:

$$\frac{\partial P(x,t)}{\partial t} = \frac{1}{2}\cdot\frac{\partial}{\partial x}\left(B(x)\cdot\frac{\partial P(x,t)}{\partial x}\right) \qquad B(x) = \lim_{{}_\Delta t\to 0}\left(\frac{\langle\langle {}_\Delta x^2\rangle\rangle}{{}_\Delta t}\right) = \frac{\langle\langle {}_\Delta x^2\rangle\rangle}{{}_\Delta t_{\min}} \qquad (11)$$

Now we may extend this model to the phase systems with phase mixing. Let us define a specific energy of a complex system:

$$\varepsilon(x,t) = \lim_{{}_\Delta t\to 0}\left(\frac{{}_\Delta x(x,t)}{{}_\Delta t}\right)^2 \approx \left(\frac{{}_\Delta x(x,t)}{{}_\Delta t_{\min}}\right)^2 \qquad (12)$$

A second transport factor can be expressed then through this quantity in the following way:

$$B(x,t) \approx \frac{\langle\langle {}_\Delta x^2\rangle\rangle}{{}_\Delta t_{\min}} = {}_\Delta t_{\min}\cdot\int \varepsilon(x,t)\cdot W(x,x_0,{}_\Delta t)dx_0 = {}_\Delta t_{\min}\cdot\langle\langle\varepsilon(x,t)\rangle\rangle \qquad (13)$$

For a conservative case an averaged specific energy is presented as: $\langle\langle\varepsilon(x,t)\rangle\rangle = \langle\langle\varepsilon(x)\rangle\rangle$. According to relation (11) this means that transport factor doesn't depend on time explicitly and standard FPK is valid. However this expression is not satisfied if $x(t) \leftrightarrow \varepsilon(t)$ mutual correspondence is absent. According to (13) the averaged specific energy conservancy may be violated in this case. A mutual correspondence is distorted because a phase trajectory mixing occurs - several system states are possible for the same phase space location. A second transport factor $B(x)$ in fact expresses diffusion in phase space, as it was remarked by G.M.Zaslavsky [6]. For the considered type of motion it has explicit time dependence: $B(x,t)$. According to the introduced division we shall consider two types of phase space behavior. First one corresponds to $\langle\langle\varepsilon(x,t)\rangle\rangle = \langle\langle\varepsilon(x)\rangle\rangle$ and can be designated as phase conservative state. Then the second one $\langle\langle\varepsilon(x,t)\rangle\rangle$ relates to a phase mixing – we shall call it transitional state. It is more general and includes phase state as the particular case.

Taking into account the non stationary properties we may modify the assumption A4 in the following way:

$$W(x,x_0,{}_\Delta t)=\delta(x-x_0)+a(x,{}_\Delta t,t)\cdot\delta'(x-x_0)+\frac{1}{2}\cdot B(x,{}_\Delta t,t)\cdot\delta''(x-x_0) \qquad (14)$$

Derivative of a given probability $P(x',t)$ is represented according to the Chapman Kolmogorov equation in the usual form:

$$\frac{\partial P(x,t)}{\partial t}=\lim_{{}_\Delta t\to 0}\left[\frac{1}{{}_\Delta t}\cdot\left(\int dx\cdot P(x_0,t)\cdot\left(W(x,x_0,{}_\Delta t)-\delta(x-x_0)\right)\right)\right] \qquad (15)$$

Substitution of (14) into (15) gives non conservative FPK:

$$\frac{\partial P(x,t)}{\partial t}=\frac{1}{2}\cdot\frac{\partial}{\partial x}\left(B(x,t)\cdot\frac{\partial P(x,t)}{\partial x}\right) \qquad (16)$$

In the given relation $D(x,t)=\sqrt{2\cdot B(x,t)}$ corresponds to an anomalous diffusion factor. We have extended the approach of Kolmogorov to an arbitrary class of systems that takes into account obligatory mixing properties.

Let's consider the simplest case of the uniform diffusion: $D(x,t)=D(t)$. This problem in fact relates to the simplest transitional case. It can be used as first approximation of chaotic description as well. A Fourier decomposition of the equation (16) then gives us following relation:

$$\int i\omega\cdot P_\omega(x,\omega)\cdot\exp(-i\omega\cdot t)d\omega=-\int B(t)\cdot(k)^2 P_k(k,t)\cdot\exp(-ik\cdot x)dk \qquad (17)$$

Here $B(t)\to B(t)/2$ is a modified transport factor. Second decomposition gives relations (18) and (19) with Fourier operator's kernel $\hat{P}(k,\omega)$:

$$P_k(k,t)=\frac{1}{\sqrt{2\pi}}\int_{-\Omega}^{\Omega}\hat{P}(k,\omega)\cdot\exp(-i\omega\cdot t)d\omega \qquad (18)$$

$$P_\omega(x,\omega)=\frac{1}{\sqrt{2\pi}}\int_{-K}^{K}\hat{P}(k,\omega)\cdot\exp(-ik_j\cdot x_j)dk_j \qquad (19)$$

Here spectral amplitudes are finite according to the normalization condition for the probability. Integrals limits are defined according to the Kotelnikov theorem: $\Omega=\left(2\cdot{}_\Delta t_{\min}\right)^{-1}, K=\left(2\cdot{}_\Delta x_{\min}\right)^{-1}$.

Substitution of (18) and (19) into equation (17) gives wave packet form of FPK:

$$\int_{-\Omega}^{\Omega}\int_{-K}^{K} i\omega\cdot\hat{P}(k,\omega)\cdot\exp(-i\omega t-ikx)dk_j d\omega=-\int_{-\Omega}^{\Omega}\int_{-K}^{K} B(t)\cdot k^2\cdot\hat{P}(k,\omega)\cdot\exp(-i\omega t-ikx)dk_j d\omega \qquad (20)$$

A general arbitrariness of integration limits allows representing of $\omega(k_i)$ dispersion law:

$$\omega(k)=i\cdot B(t)\cdot(k)^2 \qquad (21)$$

As it follows from outlined expression nonlinear dispersion law of Eq.31 is mandatory property of uniform chaotic state. Allocation of a real part leads to the relation (22):

$$\mathrm{Im}(k)=-B(t)\cdot\frac{\mathrm{Re}(k)}{\mathrm{Re}(\omega)} \qquad \mathrm{Im}(k)\le 0 \qquad (22)$$

A first Fourier decomposition of the probability density then can be given by the equation (23):

$$P(x,t)=\int P_k(k,t)\cdot\exp\left(\left|\mathrm{Im}(k)\right|\cdot x\right)\exp\left(-i\,\mathrm{Re}(k)\cdot x\right)dk \qquad (23)$$

The imaginary part of (21) corresponds to the relation (24):

$$\mathrm{Im}(\omega)=B(t)\cdot|k|^2=\alpha(t) \qquad \alpha(t)\succ 0 \qquad (24)$$

A first Fourier decomposition of the probability density can be expressed then in the following way:

$$P(x,t) = \int P_\omega(x,\omega)\cdot \exp(|\mathrm{Im}(\omega)|\cdot t)\exp(-i\,\mathrm{Re}(\omega)\cdot t)d\omega \tag{25}$$

According to the relations (23) and (25) a chaotic system with uniform dynamic properties, defined by factor $D(t)$ tends to instable clustering – small fluctuations of particles density increase by time and form attractors, repellers and stochastic islands. This process leads to destruction of dynamic uniformity and finally this approximation comes incorrect – we need to solve a general form of extended FPK (16). However it is important to notice that clustering is a natural property of chaotic systems - it relates to the first approximation of any transition. In fact we have shown how instability of phase trajectories leads to the nonlinear dispersion (21) and consequently to a space instability.

## 3 Uncertainty relations

In the previous section we have proved logically that clustering is a sufficient property of a simplest transitional state. In this section we will consider a trivial case - a phase state with a fixed boundary. However, nontrivial uncertainty relations will be derived basing on a standard one dimensional Fokker-Planck equation.
Let's consider a conservative problem of the fixed boundary:

$$\frac{\partial P(x,t)}{\partial t} = \frac{B}{2}\cdot\frac{\partial^2 P(x,t)}{\partial x^2} \tag{26}$$

$$P(0,t) = P(L,t) = 0 \qquad x\in[0,L]$$

$$P(x,0) = P_0(x)$$

The equation (26) mathematically corresponds to a uniform linear diffusion PDE. Here $B = const$ according to the condition of an energy conservancy – we consider the system's equilibrium phase state:

$$B(x,t) \approx {}_\Delta t_{\min}\cdot\varepsilon\cdot\int_0^L W(x,x_0,{}_\Delta t)dx_0 = \varepsilon\cdot{}_\Delta t_{\min} = const \tag{27}$$

A solution of (26) may be searched in a form of the Fourier expansion:

$$P(x,t) = \sum_{j=1}^{N} c_j(t)\cdot\sin\left[\left(\frac{\pi\cdot j}{L}\right)\cdot x\right] \tag{28}$$

$$c_j(t) = \frac{2}{L_x}\cdot\int_0^L P(\zeta,t)\cdot\sin\left[\left(\frac{\pi\cdot j}{L}\right)\cdot\xi\right]d\xi$$

A substitution of (28) into (26) gives the following superposition of modes:

$$\sum_{j=1}^{N}\sin\left[\left(\frac{\pi\cdot j}{L}\right)\cdot x\right]\cdot\left(\frac{\partial c_j(t)}{\partial t} + \frac{B\cdot c_j(t)}{2}\cdot\left(\frac{\pi\cdot j}{L}\right)^2\right) = 0 \tag{29}$$

For an arbitrary $L$ a second factor has an obligatorily zero value:

$$\frac{\partial c_j(t)}{\partial t} = -\frac{B\cdot c_j(t)}{2}\cdot\left(\frac{\pi\cdot j}{L}\right)^2 \tag{30}$$

Then, possible values of a transport factor $B$ may be expressed in the following way:

$$B_j = -\frac{2}{c_j(t)}\cdot\frac{dc_j(t)}{dt}\cdot\left(\frac{L}{\pi\cdot j}\right)^2 \tag{31}$$

Let's consider the instability of the particles density: $c_j(t) = c_j(0) \cdot \exp(\phi_j \cdot t)$. It leads to a clustering of particles, considered in the previous section, and finally to the non stationary regime when $B = B(t)$.

Here $c_j^0 = c_j(0)$ corresponds to an initial density probability, defined by the third relation in group (26). Factors $\phi_j$ characterize modes instability increments and depend on nonlinear properties of the complex system media. They are assumed to be constant in the given approximation for the given complex system. Here we consider the states in the vicinity of a conservative condition. A normalization condition for $P(x,t)$ can be represented in the following way:

$$\sum_{j=1}^{N} c_j^0 \cdot \exp(\phi_j \cdot t) \int_0^L \cdot \sin\left[\left(\frac{\pi \cdot j}{L}\right) \cdot x\right] dx = 1 \tag{32}$$

We can remark that satisfaction of this condition is possible only if both relations $\phi_j \succ 0$ and $\phi_j \prec 0$ are valid for the different modes of the spectrum. It means that increasing and decreasing of quasi regular fluctuations of particles density occur in the media of a complex system.

According to the relation (31) we have discrete spectrum of phases that can be realized for this considered case:

$$B_j = 2 \cdot \phi_j \cdot \left(\frac{L}{\pi \cdot j}\right)^2 \qquad \varepsilon_j = \frac{2 \cdot \phi_j}{{}_{\Delta} t_{\min}} \cdot \left(\frac{L}{\pi \cdot j}\right)^2 \qquad j = \overline{1, N} \tag{33}$$

Let's introduce a circular frequency $\omega = \dfrac{2\pi}{{}_{\Delta} t_{\min}}$. Then the relation (33) can be modified:

$$\Delta\varepsilon_j = \frac{\phi_j}{\pi^3} \cdot \left(\frac{L}{j}\right)^2 \cdot \Delta\omega \tag{34}$$

It is known that a perfect accuracy substitution for the initial coordinate of particle can be represented by Dirac function:

$$P(x(t_0)) = \delta(x(t) - x_j(t_0)) \tag{35}$$

Normalization condition for $P(x_j(t_0))$ then can be represented in the following way:

$$\int_0^L \delta(x(t) - x(t_0)) dx_j = \int_0^{L_j} \left( \sum_k \frac{\delta(t - t_0)}{\left|\partial x_j^k(t) / \partial t\right|} \right) dx \tag{36}$$

Dirac functional is represented here through the time argument.

Index $k$ corresponds to zeros of function $x(t)$. In considered case we have only one value of argument, corresponding to zero - $t_0$. Then relation (36) can be modified in the following way:

$$\lim_{t \to t_0} \left[ \int_0^L \left( \sum_k \frac{\delta(t - t_0)}{\left|\partial x^k(t) / \partial t\right|} \right) dx \right] = \frac{\partial x(t) / \partial t}{\left|\partial x(t) / \partial t\right|} \int_0^{T(L)} \delta(t - t_0) dt \tag{37}$$

As we can see in a vicinity of $t_0$ $(t \to t_0, x_j \to x_j(t_0))$ space - time correspondence allows introducing of a probability density:

$$\delta(x(t) - x(t_0)) \leftrightarrow \delta(t - t_0) \cdot sign(\partial x(t) / \partial t) \tag{38}$$

Real physical systems correspond to a finite space and time resolution: ${}_{\Delta}t \to 0$ and ${}_{\Delta}x \to 0$ in fact correspond to ${}_{\Delta}t = 2\cdot{}_{\Delta}t_{\min}$ and ${}_{\Delta}x = 2\cdot{}_{\Delta}x_{\min}$.

Limitations of our observance qualities lead to a discrete alternative of Delta function:

$$x(t) = \left[\begin{matrix} C_1 & x \prec \left|{}_{\Delta}x_{\min}\right| \\ 0 & x \geq \left|{}_{\Delta}x_{\min}\right| \end{matrix}\right. \qquad t(x_j) = \left[\begin{matrix} C_2 & t \prec \left|{}_{\Delta}t_{\min}\right| \\ 0 & t \geq \left|{}_{\Delta}t_{\min}\right| \end{matrix}\right. \tag{39}$$

Here we assume without a loss of a generality that $t_0 = 0$. According to a normalization condition for the probability, coefficients $C_1$ and $C_2$ can be expressed in the following way:

$$C_1 = \frac{1}{2\cdot{}_{\Delta}x_{\min}} \qquad C_2 = \frac{1}{2\cdot{}_{\Delta}t_{\min}} \tag{40}$$

Relations (39) correspond to a rectangular function. Its Fourier spectrum is expressed through the equations (41):

$$S_x(\omega_x) = C_1\cdot{}_{\Delta}x_{\min}\cdot \sin c\left(\frac{\omega_x\cdot{}_{\Delta}x_{\min}}{2}\right) \qquad S_t(\omega_t) = C_2\cdot{}_{\Delta}t_{\min}\cdot \sin c\left(\frac{\omega_t\cdot{}_{\Delta}t_{\min}}{2}\right) \tag{41}$$

A central maximum of the second Sinc – function is limited by the interval ${}_{\Delta}\omega_t = 2\pi/{}_{\Delta}t_{\min}$ between two first zeroes. This interval includes the biggest part of Sinc – a function integral value. In such a way ${}_{\Delta}\omega_t = 2\pi/{}_{\Delta}t_{\min}$ can be used as a characteristic width of the probability spectrum. In the same way we obtain a space frequency width:

$${}_{\Delta}\omega_t \geq \frac{2\pi}{{}_{\Delta}t_{\min}} \qquad {}_{\Delta}\omega_x \geq \frac{2\pi}{{}_{\Delta}x_{\min}} \tag{42}$$

A substitution of the first relation into (34) gives us the following expression:

$${}_{\Delta}\varepsilon_j\cdot{}_{\Delta}t_{\min} \geq \frac{2\phi_j}{\left(k_j\right)^2} \tag{43}$$

Here the wave number $k_j = \frac{\pi\cdot j}{L}$ is introduced. If we express an increment $\phi_j$ through the diffusion factor and according to the equation (33) we may derive a final uncertainty relation:

$${}_{\Delta}\varepsilon_j\cdot{}_{\Delta}t_{\min} \geq B_j \tag{54}$$

A minimal spectral diffusion $B_0 = \left(B_j\right)_{\min}$ allows us to simplify this equation in the following way:

$${}_{\Delta}\varepsilon_j\cdot{}_{\Delta}t_{\min} \geq B_0 \tag{55}$$

Let's express an energy and momentum through the coordinate in a standard way:

$$\varepsilon_j = \frac{p_j^{2}}{2} \qquad p_j = \frac{{}_{\Delta}x_j}{{}_{\Delta}t_{\min}} \tag{55}$$

Then the energy increment ${}_{\Delta}\varepsilon_j = p_j\cdot{}_{\Delta}p_j$ can be substituted in the relation (55):

$${}_{\Delta}x_j\cdot{}_{\Delta}p_j \geq B_0 \tag{56}$$

Equations (55) and (56) connect uncertainties of the dynamic state definition. If we omit the "min" and "*j*" indexes in the relation (55) these uncertainties may be represented in the following way:

$$_{\Delta}x\cdot_{\Delta}p \geq B_0 \qquad _{\Delta}\varepsilon\cdot_{\Delta}t \geq B_0 \tag{57}$$

It was shown that a fixed boundary system has a discrete number of the phase states. Each of these states is characterized by its own phase diffusion $B_j$. In such a way the diffusive spectrum depends on the selected scale $k_j = \dfrac{\pi \cdot j}{L}$. This property is well known for the case of a turbulent motion, when each scale corresponds to its own turbulent diffusion value [1]. In this case a standard molecular diffusion is only the minimal value for the molecular space scale. We have shown that for the considered case phase diffusion is the only parameter that defines a dynamic accuracy – that is why dynamic properties and a resolution in fact are scale dependent. Besides, according to the relation (33) small scales correspond to low diffusion and higher resolution abilities. Finally we come to the conclusion that the smaller scale we consider the more information we get about the motion.

## 4 Conclusions

In this paper uncertainties and clustering, lying in the basis of chaos have been investigated. It was obtained that a chaotic system with a non stationary diffusion satisfies a nonlinear dispersion law. This law leads to instabilities in a phase space and to the appearance of the clustering properties for the initially uniform system.

The generalized Fokker-Planck equation with a non stationary diffusion has been derived. It has been applied to the analysis of a fixed boundary problem. An anomalous transport turned out to be a natural property of this equation under the considered conditions. This trivial law of diffusion led to a non trivial output: relation of the coordinate – momentum uncertainties has been stated. We have shown that for the considered case phase diffusion is the only parameter that defines a dynamic accuracy – that's why dynamic properties and a resolution in fact are scale dependent.

Formulation of a generalized Fokker-Planck (GFP) equation, given in this paper, allows defining of an anomalous diffusion without complex fractal PDE. A data about the conservative spectrum of GFP may help to find stable states of real complex systems and avoid their destruction. A clustering formal explanation helps to define and anticipate attraction areas in the real phase spaces. At the same time derived uncertainty relations allows us to understand limits of the possible forecast efficiency through the connection of the accuracy with transport properties. The author believes that represented results can be used in the area of complex systems modelling and forecast.